# Brief Architectural Survey of Biopotential Recording Front-Ends since the 1970s

Taeju Lee[1] and Minkyu Je[2]

[1]Department of Electrical Engineering, Columbia University, New York, NY 10027, USA
E-mail: taeju.lee@columbia.edu

[2]School of Electrical Engineering, Korea Advanced Institute of Science and Technology, Daejeon 34141, Korea
E-mail: mkje@kaist.ac.kr

*Abstract*—Measuring the bioelectric signals is one of the key functions in wearable healthcare devices and implantable medical devices. The use of wearable healthcare devices has made continuous and immediate monitoring of personal health status possible. Implantable medical devices have played an important role throughout the fields of neuroscience, brain-machine (or brain-computer) interface, and rehabilitation technology. Over the last five decades, the bioelectric signals have been observed through a variety of biopotential recording front-ends, along with advances in semiconductor technology scaling and circuit techniques. Also, for reliable and continuous signal acquisition, the front-end architectures have evolved while maintaining low power and low noise performance. In this article, the architecture history of the biopotential recording front-ends developed since the 1970s is surveyed, and overall key circuit techniques are discussed. Depending on the bioelectric signals being measured, appropriate front-end architecture needs to be chosen, and the characteristics and challenges of each architecture are also covered in this article.

## I. INTRODUCTION

SINCE the 1970s, the biopotential recording front-ends have evolved based on various circuit architectures to measure bioelectric signals, leading to significant advancements in wearable healthcare devices (WHDs) and implantable medical devices (IMDs). The WHDs have been widely used for daily personal healthcare and continuous patient monitoring. The IMDs have been used in patients for a variety of purposes, such as deep brain stimulation for Parkinson's, artificial pulse generation for heart failure, stimulation of auditory nerve for hearing loss, etc. The recording front-ends have made a significant achievement in the brain-machine interface, revolutionizing the motor function restoration of patients with difficulties in their physical activities. In addition, the recording front-ends have significantly advanced the field of neuroscience by enabling *in-vivo* and *in-vitro* neural activity monitoring.

In this article, the recording front-ends, developed since the 1970s, are surveyed regarding their front-end architectures and key circuit techniques. Note that the front-ends that record bioelectric signals invasively and non-invasively have been surveyed, and the bioelectric signals include action potentials (APs), local field potentials (LFPs), electrocorticogram, electroencephalogram, electrocardiogram, electromyogram, and electrooculogram.

Over the last decades, the complementary metal-oxide semiconductor (CMOS) devices have been gradually scaled down, and the recording front-ends also have evolved along with device scaling, as shown in Fig. 1. The front-ends have had a significant impact across a variety of fields, from personal healthcare to cutting-edge neuroscience research, and the development of the front-end is actively underway to improve the performance such as spatial density, power consumption, noise, etc. The biopotential recording front-ends have been developed based on various circuit architectures, and each architecture shows different characteristics in its frequency response, DC offset cancellation, input-referred noise, etc. Also, the circuit technique can be combined with the circuit architecture to improve a specific performance, e.g., the input-referred noise, input impedance, and dynamic range. However, the circuit techniques may introduce performance trade-offs. Therefore, choosing appropriate architecture and circuit techniques depending on the goal is important to minimize performance limitations and trade-offs. All front-ends surveyed in this article are listed in the paper [1].

Section II of this article provides a chronological explanation of circuit architectures that emerged as technology scaled. Section III presents the key front-end architectures used over the past decades. Section IV concludes this article by outlining future directions.

## II. ARCHITECTURAL PROGRESS & TECHNOLOGY SCALING

### A. 1970s to 1980s

Advances in microfabrication technology triggered the miniaturization of front-ends, which led to the development of implantable devices. In the early development period, the recording front-ends were designed as a continuous-time buffer (CT-Buf) using the source follower [2]–[5]. In 1986, a front-end architecture based on continuous-time open-loop amplification (CT-OL-Amp) was designed using a 6-$\mu$m NMOS process [6]. In 1987, a front-end using current balancing amplification was developed in a 3-$\mu$m CMOS process [7]. Note that Fig. 1 shows the architectural progress as technology has scaled down since the 1970s, which is explored using the technology scaling data in [1].

### B. 1990s

Although many recording front-ends were not developed in the 1990s, the CT-OL-Amp-based front-ends were developed using the 6-$\mu$m and 3-$\mu$m nodes in 1991 and 1992, respectively, as shown in Fig. 1. Also, a current balancing-based front-end was developed using a 2.4-$\mu$m CMOS process [8], and a resistive-feedback-based channel was developed using a 2-$\mu$m CMOS process [9]. Most front-ends implemented from the 1970s to the 1990s were developed to detect neural activities such as APs and LFPs. Also, most front-ends developed in that period were designed as active neural probes that placed the front-end close to the electrode, thereby minimizing the form factor of the recording system and improving the observed signal quality.

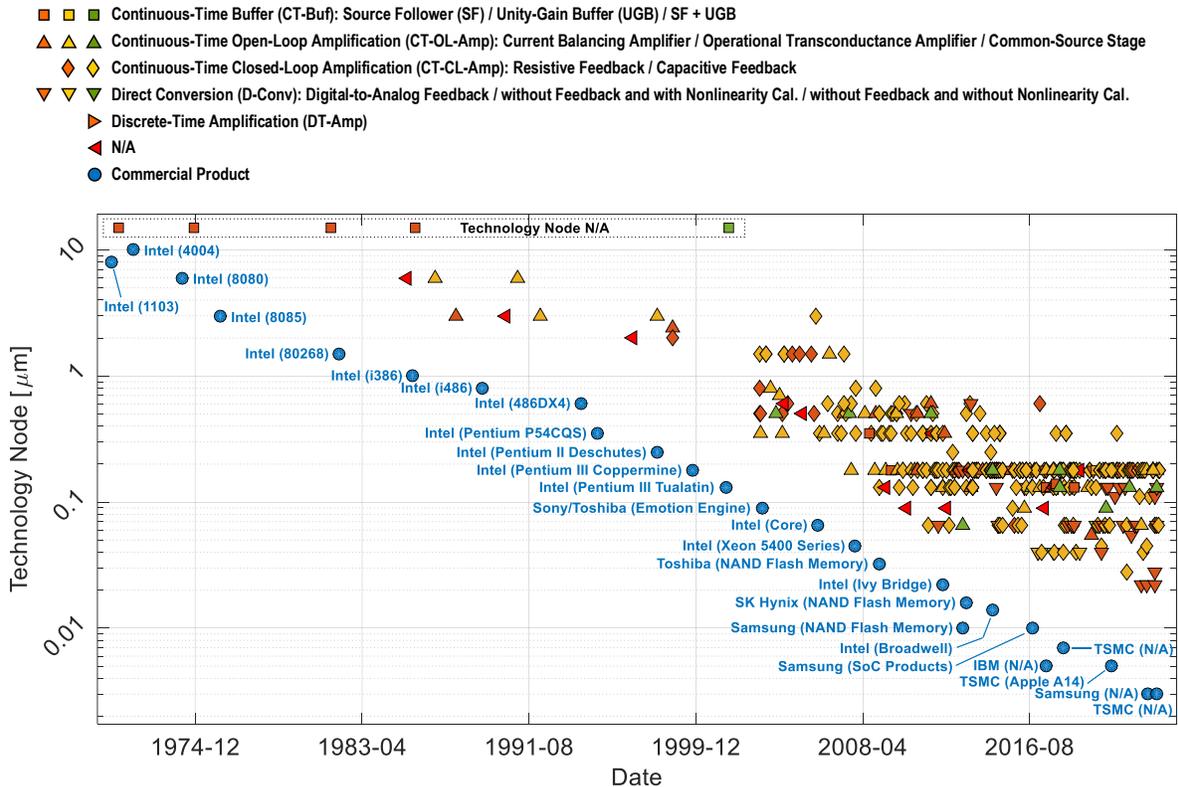

Fig. 1. Architecture trends of biopotential recording front-ends depending on technology scaling.

*C. 2000s*

Compared to the period from the 1970s to the 1990s, numerous front-ends were developed using various technology nodes ranging from 1.5 $\mu$m to 0.18 $\mu$m in the 2000s, as shown in Fig. 1. Especially, after the emergence of the capacitive-feedback-based architecture that efficiently filters out input DC offsets and exhibits excellent noise performance [10], many front-ends were developed by employing that architecture. Although resistive-feedback-based front-ends were also widely employed in the early 2000s, the capacitive-feedback architecture was overwhelmingly used in front-end design throughout that period. During this period, the front-ends based on CT-OL-Amp architectures were rarely developed compared to the continuous-time closed-loop amplification (CT-CL-Amp) architectures.

*D. 2010s*

Even into the 2010s, among the closed-loop front-end architectures employing resistive feedback and capacitive feedback, the capacitive-feedback-based one was dominantly adopted in front-end designs. Note that 1) the CT-Buf-based front-end means that the front-end is designed as a source follower or a unity-gain buffer; 2) the CT-OL-Amp-based front-end means that the front-end is implemented using one of the current balancing amplifier, operational transconductance amplifier (OTA), and common-source stage; and 3) the CT-CL-Amp-based front-end means that the front-end is designed based on resistive feedback or capacitive feedback.

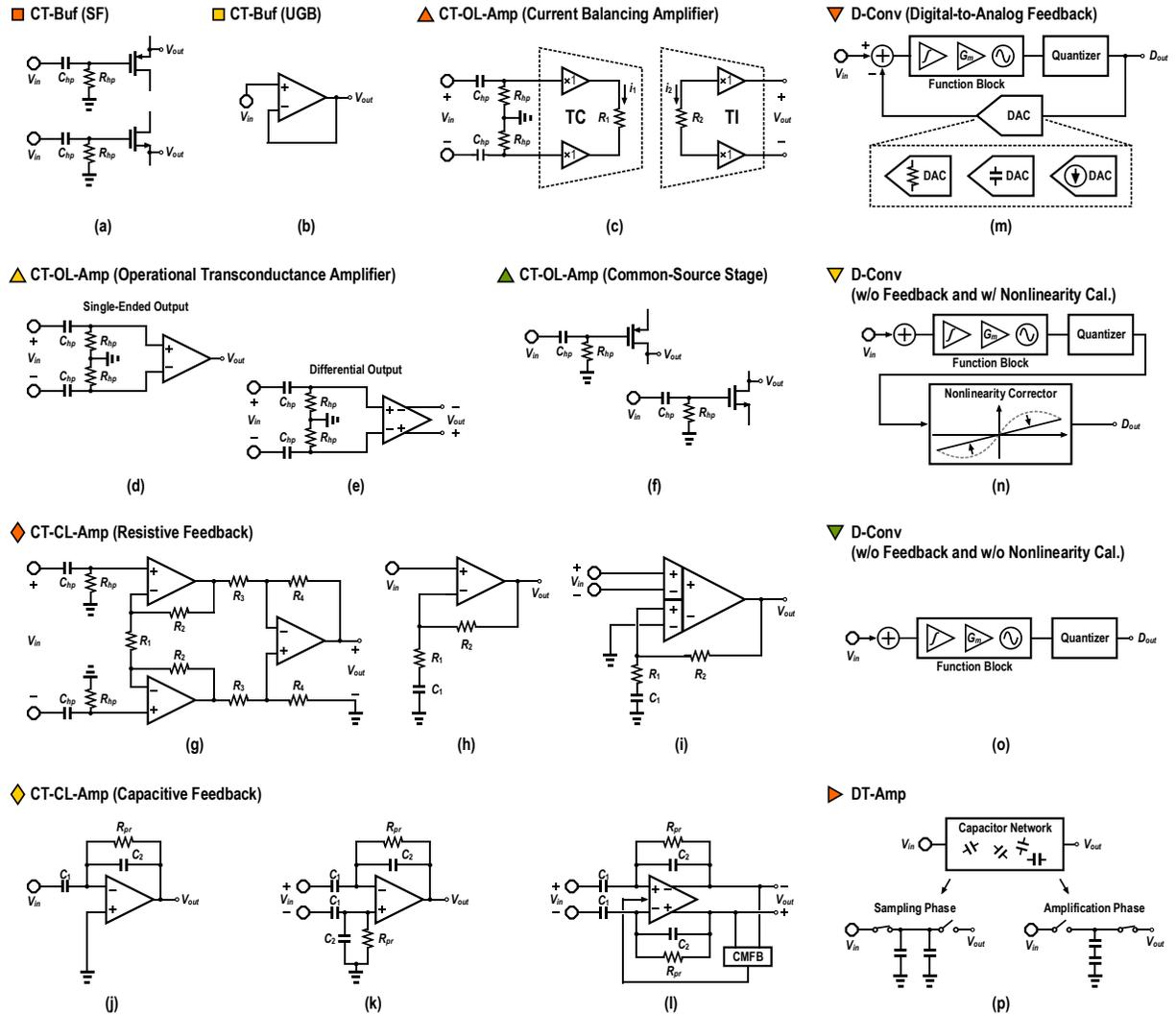

Fig. 2. Overall front-end architectures used over the past decades.

The architectures based on CT-OL-Amp and CT-CL-Amp require an analog-to-digital converter (ADC) following the analog front-end stage to digitize the signal. These front-end architectures could induce performance degradation in their dynamic range, spatial density, etc. In the early 2010s, a direct conversion (D-Conv) architecture, which directly generates the digitized data from the input, emerged [11]–[13]. The D-Conv architecture began to be widely used to compensate for the drawbacks of previous architectures employing CT-OL-Amp and CT-CL-Amp. Also, the advent of a discrete-time amplification (DT-Amp) architecture led to the improvement of the noise efficiency factor (NEF) of the front-end compared to the previous front-end using an OTA [14]. During this period, the front-ends were developed using the technology nodes ranging from 0.6 $\mu$m to 40 nm, as shown in Fig. 1.

*E. 2020s*

In the 2020s, the architectures based on D-Conv and CT-CL-Amp using capacitive feedback have been more widely used for front-end design than other topologies. Also, the NEF and power efficiency factor (PEF) have been further improved in the DT-Amp architecture [15]. The front-ends have been developed using increasingly advanced technology nodes, as shown in Fig. 1. The technology nodes used have ranged from 0.35 $\mu$m to 22 nm. Considering that the D-Conv architecture can be designed in a digitally intensive manner compared to CT-OL-Amp and CT-CL-Amp architectures, hopefully, the front-end performance, such as power and area consumption, could be continuously improved in the future, especially when using more advanced technology nodes.

### III. FRONT-END ARCHITECTURE

Fig. 2 summarizes the front-end architectures for biopotential recording developed over the last five decades. First, the front-ends are categorized into three groups: continuous-time amplification (CT-Amp), direct conversion (D-Conv), and discrete-time amplification (DT-Amp). Second, the CT-Amp architectures are categorized into two groups: continuous-time closed-loop amplification (CT-CL-Amp) and continuous-time open-loop amplification (CT-OL-Amp). Third, the CT-CL-Amp architectures are again divided into two groups according to their feedback types: resistive feedback (Figs. 2(g)–(i)) and capacitive feedback (Figs. 2(j)–(l)). Fourth, the CT-OL-Amp architectures are divided into three groups according to their amplifier types: the current balancing amplifier (Fig. 2(c)), OTA (Figs. 2(d) and (e)), and common-source stage (Fig. 2(f)). Fifth, the D-Conv architectures are categorized according to whether their nonlinearities are compensated for by using feedback or calibration (Figs. 2(m)–(o)). Finally, the CT-Buf architectures are divided into two types using the source follower and unity-gain buffer, shown in Figs. 2(a) and (b), respectively. Fig. 3 visualizes the architectures used over the last decades chronologically, revealing how frequently each architecture has been used over time.

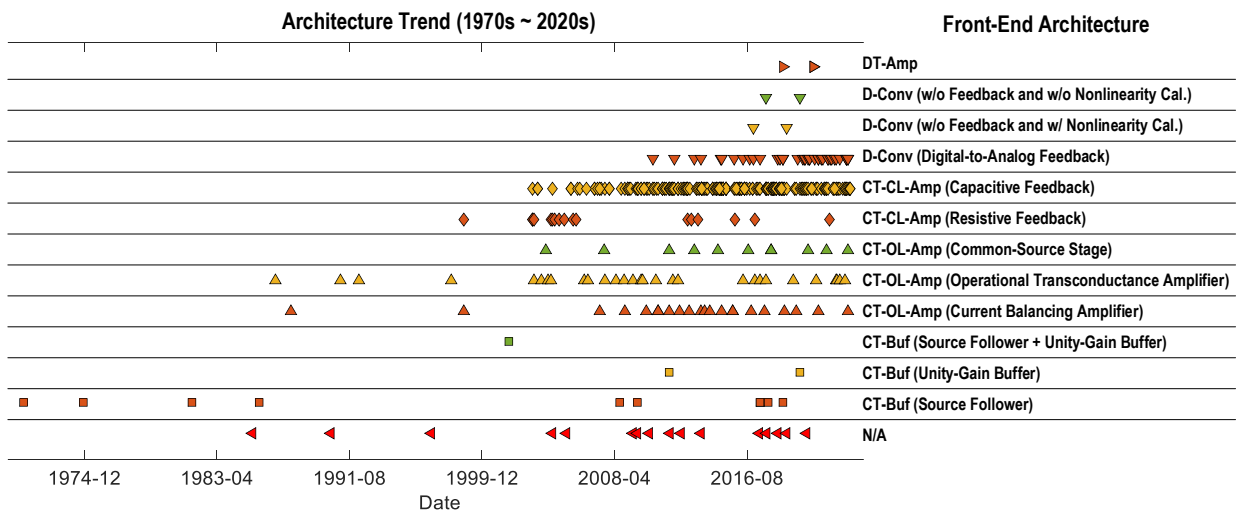

Fig. 3. Front-end architectures used in chronological order.

## IV. CONCLUSION

This article presents a history of biopotential recording front-ends. The recording front-ends are chronologically surveyed from the viewpoints of technology scaling and architectural development. As CMOS technology scales down, the front-ends also have evolved, using more advanced nodes to improve spatial density and power efficiency. To record from the significant portion of the brain network, which could be the most challenging task, scalable front-ends with higher spatial density and energy efficiency must be realized without compromising other performance parameters such as noise and dynamic range. To achieve this goal, the revolutionization needs to continue in device scaling, front-end architectures, circuit techniques, and neural probes.

## REFERENCES

[1] T. Lee and M. Je, "Trend investigation of biopotential recording front-end channels for invasive and non-invasive applications," *arXiv preprint arXiv:2305.13463*, pp.1–31, May 2023.

[2] K. D. Wise and J. B. Angell, "A microprobe with integrated amplifiers for neurophysiology," in *Proc. IEEE Int. Solid-State Circuits Conf. Dig. Tech. Papers*, Feb. 1971, pp. 100–101.

[3] T. Matsuo and K. D. Wise, "An integrated field-effect electrode for biopotential recording," *IEEE Trans. Biomed. Eng.*, vol. BME-21, no. 6, pp. 485–487, Nov. 1974.

[4] D. T. Jobling, J. G. Smith, and H. V. Wheal, "Active microelectrode array to record from the mammalian central nervous system *in vitro*," *Med. Biol. Eng. Comput.*, vol. 19, pp. 553–560, Sept. 1981.

[5] M. G. Dorman, M. A. Prisbe, and J. D. Meindl, "A monolithic signal processor for a neurophysiological telemetry system," *IEEE J. Solid-State Circuits*, vol. SC-20, no. 6, pp. 1185–1193, Dec. 1985.

[6] K. Najafi and K. D. Wise, "An implantable multielectrode array with on-chip signal processing," *IEEE J. Solid-State Circuits*, vol. SC-21, no. 6, pp. 1035–1044, Dec. 1986.

[7] M. S. J. Steyaert, W. M. C. Sansen, and C. Zhongyuan, "A micropower low-noise monolithic instrumentation amplifier for medical purposes," *IEEE J. Solid-State Circuits*, vol. SC-22, no. 6, pp. 1163–1168, Dec. 1987.

[8] R. Martins, S. Selberherr, and F. A. Vaz, "A CMOS IC for portable EEG acquisition systems," *IEEE Trans. Instrum. Meas.*, vol. 47, no. 5, pp. 1191–1196, Oct. 1998.

[9] J. J. Pancrazio *et al.*, "Description and demodulation of a CMOS amplifier-based-system with measurement and stimulation capability for bioelectrical signal transduction," *Biosens. Bioelectron.*, vol. 13, no. 9, pp. 971–979, Oct. 1998.

[10] R. R. Harrison and C. Charles, "A low-power low-noise CMOS amplifier for neural recording applications," *IEEE J. Solid-State Circuits*, vol. 38, no. 6, pp. 958–965, Jun. 2003.

[11] Y. M. Chi and G. Cauwenberghs, "Micropower integrated bioamplifier and auto-ranging ADC for wireless and implantable medical instrumentation," in *Proc. IEEE European Solid-State Circuits Conf.*, Sept. 2010, pp. 334–337.

[12] R. Muller, S. Gambini, and J. M. Rabaey, "A 0.013 mm$^2$, 5 $\mu$W, DC-coupled neural signal acquisition IC with 0.5 V supply," *IEEE J. Solid-State Circuits*, vol. 47, no. 1, pp. 232–243, Jan. 2012.

[13] Y. Li, D. Zhao, and W. A. Serdijn, "A sub-microwatt asynchronous level-crossing ADC for biomedical applications," *IEEE Trans. Biomed. Circuits Syst.*, vol. 7, no. 2, pp. 149–157, Apr. 2013.

[14] T. Jang *et al.*, "A noise-efficient neural recording amplifier using discrete-time parametric amplification," *IEEE Solid-State Circuits Lett.*, vol. 1, no. 11, pp. 203–206, Nov. 2018.

[15] G. Atzeni, A. Novello, G. Cristiano, J. Liao, and T. Jang, "A 0.45/0.2-NEF/PEF 12-nV/√Hz highly configurable discrete-time low-noise amplifier," *IEEE Solid-State Circuits Lett.*, vol. 3, pp. 486–489, Oct. 2020.

**Taeju Lee** received the B.S. degree in Electrical, Electronics and Communication Engineering from Korea University of Technology and Education (KOREATECH), Cheonan-si, South Korea, in 2014, the M.S. degree in Information and Communication Engineering from Daegu Gyeongbuk Institute of Science and Technology (DGIST), Daegu, South Korea, in 2016, and the Ph.D. degree in Electrical Engineering from Korea Advanced Institute of Science and Technology (KAIST), Daejeon, South Korea, in 2021.

From 2021 to 2022, he was a Postdoctoral Researcher at KAIST, Daejeon, Korea. Since 2022, he has been a Postdoctoral Research Scientist at Columbia University, New York, NY, USA. His current research interests include integrated circuits used in neuromodulation technology and ultrasound systems.

**Minkyu Je** received the M.S. and Ph.D. degrees, both in Electrical Engineering and Computer Science, from Korea Advanced Institute of Science and Technology (KAIST), Daejeon, Korea, in 1998 and 2003, respectively. In 2003, he joined Samsung Electronics, Giheung, Korea, as a Senior Engineer. From 2006 to 2013, he was with Institute of Microelectronics (IME), Agency for Science, Technology and Research (A*STAR), Singapore. From 2011 to 2013, he led the Integrated Circuits and Systems Laboratory at IME as a Department Head. He was also a Program Director of NeuroDevices Program under A*STAR Science and Engineering Research Council (SERC) from 2011 to 2013, and an Adjunct Assistant Professor in the Department of Electrical and Computer Engineering at National University of Singapore (NUS) from 2010 to 2013. He was an Associate Professor in the Department of Information and Communication Engineering at Daegu Gyeongbuk Institute of Science and Technology (DGIST), Korea from 2014 to 2015. Since 2016, he has been an Associate Professor in the School of Electrical Engineering at Korea Advanced Institute of Science and Technology (KAIST), Korea.

His main research areas are advanced IC platform development including smart sensor interface ICs, ultra-low-power wireless communication ICs, high-efficiency energy supply and management ICs, ultra-low-power timing ICs, resource-constrained computing ICs, as well as microsystem integration leveraging the advanced IC platform for emerging applications such as intelligent miniature biomedical devices, ubiquitous wireless sensor nodes, and future mobile devices. He is an editor of 1 book, an author of 6 book chapters, and has more than 360 peer-reviewed international conference and journal publications. He also has more than 70 patents issued or filed. He has served on the Technical Program Committee and Organizing Committee for various international conferences, symposiums, and workshops including IEEE International Solid-State Circuits Conference, IEEE Asian Solid-State Circuits Conference, and IEEE Symposium on VLSI Circuits. He has also been a Distinguished Lecturer of IEEE Circuits and Systems Society from 2020 to 2022.